\documentclass[12pt,preprint]{aastex}

\shorttitle{Confined flare} \shortauthors{Li et al.}

\begin{document}

\title{Three-dimensional Magnetic Reconnection Triggering an X-class
Confined Flare in Active Region 12192}

\author{Ting Li\altaffilmark{1,2}, Yijun Hou\altaffilmark{1,2}, Shuhong Yang\altaffilmark{1,2} \& Jun Zhang\altaffilmark{1,2}}

\altaffiltext{1}{CAS Key Laboratory of Solar Activity, National
Astronomical Observatories, Chinese Academy of Sciences, Beijing
100101, China; liting@nao.cas.cn} \altaffiltext{2}{School of
Astronomy and Space Science, University of Chinese Academy of
Sciences, Beijing 100049, China}

\begin{abstract}

We present an extensive analysis of the X2.0-class confined flare on
2014 October 27 in the great active region AR 12192, observed by the
\emph{Interface Region Imaging Spectrograph} and the \emph{Solar
Dynamics Observatory}. The slipping motion of the substructures
within the negative-polarity flare ribbon (R1) and continual
reconnection-induced flows during the confined flare are first
presented. The substructures within ribbon R1 were observed to slip
in opposite directions at apparent speeds of 10-70 km s$^{-1}$. The
slipping motion exhibited the quasi-periodic pattern with a period
of 80-110 s, which can be observed since the flare start and
throughout the impulsive phase of the flare. Simultaneously
quasi-periodic flows moved along a reverse-S shaped filament, with
an average period of about 90 s. The period of reconnection-induced
flows is similar to that of the slippage of ribbon substructures,
implying the occurrence of quasi-periodic slipping magnetic
reconnection. The spectral observations showed that the Si {\sc iv}
line was blueshifted by 50-240 km s$^{-1}$ at the location of the
flows. During the process of the flare, the filament did not show
the rise phase and was not associated with any failed eruption. The
flare mainly consisted of two sets of magnetic systems, with both of
their east ends anchoring in ribbon R1. We suggest that the slipping
magnetic reconnection between two magnetic systems triggers the
confined flare.

\end{abstract}


\section{Introduction}

Solar flares are explosive manifestations of magnetic energy in the
solar corona (Forbes et al. 2006). The eruptive flares are
associated with coronal mass ejections (CMEs) that constitute major
drivers for space weather (Svestka \& Cliver 1992). The confined
flares are not accompanied by CMEs, however, they are important to
reveal the physical nature of solar flares and the relationship
between flares and CMEs.

Many previous studies have attempted to explain why a flare develops
into a confined event (Gorbachev \& Somov 1989; Schmieder et al.
1997; Nindos \& Andrews 2004; Yang et al. 2014; Dalmasse et al.
2015; Jing et al. 2015; Liu et al. 2014). Among the reported
confined flares, one category is related to loop-loop interactions,
which will not lead to CMEs due to the absence of a filament or flux
rope as the seed of a CME (Green et al. 2002; Sui et al. 2006; Ning
et al. 2018). This kind of confined flares often shows brightenings
in compact loop system, and has short time scales (e.g., on the
order of a few minutes) and small size scales (Pallavicini et al.
1977). The other category of confined flares is attributed to high
strength and low decay index of the background magnetic field that
prevent magnetic energy release (Fan \& Gibson 2007; Cheng et al.
2011; Zuccarello et al. 2017), and either the filament fails to
erupt (Ji et al. 2003; Guo et al. 2012; Kushwaha et al. 2015) or the
filament successfully erupts and no material is ejected into the
interplanetary space (Yang \& Zhang 2018; Li et al. 2018).
T{\"o}r{\"o}k \& Kliem (2005) simulated a confined eruption driven
by the kink instability of a coronal magnetic flux rope, and
suggested that the gradual decrease of the overlying field with
height was a main cause of confined flares. Wang \& Zhang (2007)
statistically investigated the X-class flares occurring during
1996$-$2004 and found that the confined flares (about 10\%) occurred
closer to the magnetic center, which implied that a stronger
overlying constraining field may determine whether a flare is
confined.

Magnetic reconnection is believed to be the fundamental process that
converts the stored magnetic energy into kinetic energy of
accelerated particles and plasma and thermal energy during a flare
(Priest \& Forbes 2002). In the classical 2D picture, magnetic
reconnection can occur at null points, where the magnetic field
vanishes (Lau \& Finn 1990; Wang 1997). In 3D, the quasi-separatrix
layers (QSLs; D{\'e}moulin et al. 1996), i.e., thin layers
characterized by very strong magnetic connectivity gradients, are
also preferred sites for current accumulation and energy dissipation
(Mandrini et al. 1997; Pontin et al. 2016). In the 3D reconnecting
mode, magnetic field lines ``slip" or ``flip" inside the plasma due
to the continuous exchange of magnetic connectivity, and thus the
magnetic reconnection occurring in the QSLs is referred as slipping
magnetic reconnection if the speed of slipping motion is
sub-Alfv\'{e}nic (Priest \& D{\'e}moulin 1995; Aulanier et al.
2006). The footprints of the QSLs have been shown to coincide with
the observed flare ribbons (D{\'e}moulin et al. 1997; Dalmasse et
al. 2015; Savcheva et al. 2015). The motions of ultraviolet (UV)
brightening along the ribbons are considered as important signatures
of slipping magnetic reconnection along the QSLs (Aulanier et al.
2006; Janvier et al. 2013). In recent years, several studies have
found the evidence of slipping magnetic reconnection during eruptive
flares (Dud{\'{\i}}k et al. 2014, 2016; Li \& Zhang 2014, 2015;
Zheng et al. 2016; Jing et al. 2017). Li et al. (2015) showed the
quasi-periodic slipping motion of bright kernels within flare
ribbons, suggestive of the occurrence of quasi-periodic slipping
magnetic reconnection during an eruptive flare. Dud{\'{\i}}k et al.
(2016) presented that the flare loops slipped in opposite directions
along flare ribbons, consistent with the prediction of the standard
solar flare model in 3D (Aulanier et al. 2012; Janvier et al. 2014).

Until now, the detailed signatures of slipping magnetic
reconnections have been presented only in eruptive flares. How about
the confined flares? Can the slipping magnetic reconnection trigger
the confined flares? Why does the flare develop into a confined
event? In this paper, based on the observations from the
\emph{Interface Region Imaging Spectrograph} (\emph{IRIS}; De
Pontieu et al. 2014) and the \emph{Solar Dynamics Observatory}
(\emph{SDO}; Pesnell et al. 2012), we investigate in detail the
features of the 3D reconnection process during a confined X2.0 flare
on 2014 October 27 in the great solar active region (AR) 12192.
Especially, the slipping motion of the substructures within a flare
ribbon and continual reconnection-induced flows during the confined
flare are first presented.

NOAA AR 12192 was the biggest sunspot region in solar cycle 24, and
it produced 6 X-class flares, 22 M-class flares, and 53 C-class
flares during its disk passage. Interestingly, all the X-class
flares in the AR are confined flares and not associated with any
CMEs. Previous studies showed that AR 12192 exhibited weaker
non-potentiality and stronger confinement from the overlying
magnetic field, which determine the poor CME productivity of the AR
(Sun et al. 2015; Chen et al. 2015; Liu et al. 2016; Sarkar \&
Srivastava 2018). However, Zhang et al. (2017) suggested that the
complexity of the involved magnetic field structures may be
responsible for the confined nature. The strongest X3.1-class flare
on 2014 October 24 has been extensively studied by using nonlinear
force-free field (NLFFF) extrapolations and numerical simulations.
Inoue et al. (2016) and Jiang et al. (2016) suggested that the AR
was composed of multiple weak-twisted flux tubes and the
tether-cutting reconnection between these sheared arcades triggered
the X3.1-class flare. Prasad et al. (2018) attributed the flare
onset to the null-point reconnections in a fan-spine topology, and
they found no flux rope was involved in the flare. Contrarily, a
magnetic flux rope was present in the numerical simulation of Amari
et al. (2018), who suggested that a strong magnetic cage of
overlying field could explain the confined nature.

The structure of the paper is as follows. In Section 2, we describe
the \emph{IRIS} and \emph{SDO} observations and the methods of data
analysis. Section 3 presents the dynamic evolution of flare loops
and ribbons, and the flows along a non-eruptive filament. Finally,
in Section 4, we discuss and summarize our work.

\section{Observations and Data Analysis}

In this study, we focus on the last X-class flare in AR 12192
occurring near the west limb (S$18^{\circ}$, W$57^{\circ}$) on 2014
October 27. The event has been analyzed by Polito et al. (2016), who
investigated the dynamics of the chromospheric plasma at the flare
kernels. The soft X-ray light curve from the Geostationary
Operational Environment Satellite (GOES) 1$-$8 {\AA} channel shows
that the X2.0-class flare started at 14:12 UT, peaked at 14:47 UT
and ended at 15:09 UT. The \emph{IRIS} performed an eight-step
coarse raster observation from 14:04 UT to 17:44 UT, with the step
of 2$\arcsec$, giving a total field of view (FOV) of
14$\arcsec$$\times$ 119$\arcsec$ for the spectra. The step cadence
of the spectral observation was $\sim$3.2 s and each raster lasted
for about 26 s. The \emph{IRIS} slit jaw images (SJIs) were taken at
a cadence of 13 s in the wavelengths 1330 {\AA} (at steps 0, 4),
2796 {\AA} (at steps 1, 5) and 2832 {\AA} (at steps 3, 7). The
spatial pixel size of the images is $\sim$0$\arcsec$.332. We mainly
use the 1330 {\AA} SJIs to investigate the detailed evolution of
flare ribbons. The 1330 {\AA} filter records emission mainly from
the UV continuum and two C {\sc ii} lines, which are formed in the
lower transition region (TR) with a temperature of about 25000 K
(Tian et al. 2014). The spectral observations of Si {\sc iv} 1402.77
{\AA} window are mainly analyzed in our study. The Si {\sc iv}
1402.77 {\AA} profile samples gas at $\sim$80000 K equilibrium
temperatures in the middle TR (Peter et al. 2014). We apply the
single-Gaussian fit to obtain the Doppler shift and the line width
(1/\emph{e} width). The non-thermal width is also calibrated by
assuming the thermal broadening of 6.88 km s$^{-1}$ (T$_{e}$=80000 K
of Si {\sc iv} line) and the instrumental broadening of 4.1 km
s$^{-1}$ (De Pontieu et al. 2014).

We also use the E(UV) images from the Atmospheric Imaging Assembly
(AIA; Lemen et al. 2012) and line-of-sight (LOS) magnetograms from
the Helioseismic and Magnetic Imager (HMI; Scherrer et al. 2012) on
board the \emph{SDO}. The AIA instrument observes the Sun in 10
UV/EUV filters with the cadences of 24$/$12 s and the pixel size of
0$\arcsec$.6. We focus on several different filters including 1600,
304, 171 and 131 {\AA} to fully understand the dynamic evolution of
the flare and a non-eruptive filament during the flare. The HMI LOS
magnetograms are used to reveal the magnetic structures where flare
ribbons are located. We co-align the AIA images and \emph{IRIS} 1330
{\AA} SJIs by identifying the characteristic features.

%The neutral line S {\sc i} 1401.5136 {\AA} is used for the
%wavelength calibration of the Si {\sc iv} 1402.77 {\AA} spectral
%window.

\section{Results}

\subsection{Overview of the Event}

As shown from the high-temperature 131 {\AA} observations (about 11
MK; O'Dwyer et al. 2010), the flare loops displayed a clear ``X
shape" in the initial stage of the flare (Figures 1(a) and (b); also
see the animation of Figure 1). Two curved flare ribbons R1 and R2
were formed as seen in 1600 {\AA} images (Figure 1(d)). The east
footpoints of the X-shaped flare loops were located at ribbon R1 and
their west footpoints anchored in ribbon R2. Then the east
footpoints of the ``X-shaped" structure S1 elongated northward and
meanwhile another large-scale structure S2 appeared, connecting
ribbon R1 and a new ribbon R3 (Figures 1(b) and (e)). Afterwards,
the east footpoints (R1) of the structure S2 slipped towards the
south while the west footpoints (R3) slipped in the opposite
direction (red arrows in Figure 1(c)). The comparison of 1600 {\AA}
images with HMI LOS magnetograms showed that ribbon R1 was located
at the negative-polarity sunspot, ribbon R2 at the leading
positive-polarity fields of the AR, and ribbon R3 anchoring at the
peripheral positive-polarity fields (Figure 1(f)).

\subsection{Dynamics of Flare Loops and Flare Ribbons}

The 171 {\AA} and 304 {\AA} observations showed that a ``reverse
S-shaped" filament existed between the two flare ribbons R1 and R2
(Figures 2(a)-(b); also see the animation of Figure 2), with a
length of about 70 Mm. The initial brightenings at the east part of
the filament appeared from about 8 min after the flare onset (14:20
UT; see Figure 2(c)), indicating that the reconnection during the
flare resulted in the brightenings and activation of the filament.
Then the entire filament was brightened and seemed to undergo a tiny
torsional deformation (Figure 2(d)). Continual mass flows were
observed from the middle part of the filament to the east end
(Figure 2(d)), implying the occurrence of magnetic reconnection
between the filament and the nearby magnetic structures. The
filament showed no evident rise during the whole event and still
existed at the end of the flare.

The detailed evolution of flare ribbons can be obtained from the
\emph{IRIS} 1330 {\AA} observations. The ribbon R1 showed an
elongation motion towards the north almost from the flare start (red
arrows in Figures 2(e)-(g)). Meanwhile, the bright substructures
within ribbon R1 exhibited a slipping motion in the opposite
direction (brown circles and arrows in Figures 2(f)-(g)). Associated
with the elongation of the negative-polarity ribbon R1, the
positive-polarity ribbon R2 also showed an elongation motion towards
the south (blue arrows in Figures 2(e)-(g)). Simultaneously, the
separation between ribbons R1 and R2 increased as they moved away
from the polarity inversion line. As ribbon R1 reached the eastern
footpoints of the filament, the filament material was illuminated in
the 1330 {\AA} image (Figure 2(g)) and continual
reconnection-induced flows were detected along the non-eruptive
filament (Figure 2(h)). The high-temperature flare loops connecting
the two ribbons R1 and R2 were traced out in the 131 {\AA} images
(Figures 2(i)-(k)). Two flare loops L1 and L2 comprised the ``X
shape" in the initial stage (Figure 2(i)), with their east end at
ribbon R1 and west end at ribbon R2. Associated with the elongation
motions of ribbons R1 and R2 in opposite directions, more flare
loops (L3 and L4 in Figures 2(j)-(k)) were traced out in the 131
{\AA} observations. These flare loops at different times (Figure
2(k)) delineated a ``bi-fan" surface (Li \& Zhang 2014), implying
the complicated magnetic structure involved in the flare (Polito et
al. 2016). The comparison of the 1330 {\AA} and 131 {\AA} images
shows that the 1330 {\AA} substructures within the evolving ribbon
are nearly cospatial with the footpoints of the 131 {\AA} loops. The
relationship between the flare kernels observed in 1600 {\AA} or
1700 {\AA} and the high-temperature flare loops has already been
reported in Dud{\'{\i}}k et al. (2014, 2016). The emission intensity
of the flare loops was enhanced significantly as the flare
developed, and the reconnection-induced flows seemed to originate
from the highly sheared location (Figure 2(l)).

The 1330 {\AA} observations showed that the east ribbon R1 was
composed of numerous dot-like substructures, which slipped in
opposite directions towards both ends of the ribbon (Figure 3 and
the associated animation). We followed the trails of 5 different
substructures within ribbon R1 and displayed them in Figure 3.
Almost from the flare start, bright dot ``1" showed the apparent
slipping motion towards the straight part (SP in panel (g)) of the
ribbon, with a displacement of about 12 Mm in 3 min and an average
speed of 70 km s$^{-1}$ (panels (a)-(c)). However, bright dots ``2"
and ``3" slipped towards the hook of the ribbon at a slower speed of
about 20 km s$^{-1}$ (panels (d)-(f)). Similarly, bright dots ``4"
and ``5" moved in opposite directions (panels (g)-(i)), and the
bi-directional slippage can be observed until the flare peak time
(panel (i)).

In order to analyze the slipping motions of the substructures, we
obtain the stack plot (Figure 4(a)) along slice ``A$-$B" in 1330
{\AA} images (red curve in Figure 3(g)). As seen from the stack
plot, multiple bright stripes were detected, with each stripe
representing the apparent slipping motion of one single
substructure. The speed of the slipping motion was about 20-30 km
s$^{-1}$ in the first stage before 14:35 UT. Then the substructures
within ribbon R1 were illuminated again, implying the occurrence of
two-step energy release in the event. In the second stage after
14:35 UT, the majority of the substructures exhibited the slipping
motion from ``B" to ``A" and a small fraction was seen to slip in
the opposite direction, with apparent speeds of 10-40 km s$^{-1}$.
Along two horizontal slices (``H1" and ``H2" in Figure 4(a)), we
also obtain the intensity-time profiles (black and red curves in
Figure 4(b)) at two certain locations. The two profiles both showed
the quasi-periodic pattern, indicating that at the same location the
substructure intermittently appeared and passed by. The average
period of the slipping motion was about 110 s in the first stage and
80 s in the second episode. The slipping motion lasted about 30 min
and was observed throughout the impulsive phase of the flare (Figure
4(b)). The spatial size of the substructures was estimated as about
500 km (Figures 4(c)-(d)), comparable to the resolution limit of the
\emph{IRIS} instrument of about 480 km. The substructures observed
in 1330 {\AA} are larger than the fine structures within H$\alpha$
flare ribbons in another event ($\sim$100 km in Sharykin \&
Kosovichev 2014). The difference is probably caused by the low
spatial resolution of the \emph{IRIS} observations compared to the
high-resolution H$\alpha$ images.

\subsection{Reconnection-induced Flows Along the Non-eruptive Filament}

About 8 min after the flare onset ($\sim$14:20 UT), the reverse-S
shaped filament was activated and continual mass flows appeared from
its middle part to the east end. Along the main axis of the
filament, we obtained two stack plots (Figures 5(c)-(d))
respectively along slice ``C-D" in 1330 {\AA} images (red curve in
Figure 5(a)) and slice ``E-F" in 304 {\AA} images (blue curve in
Figure 5(b)). As seen from the 1330 {\AA} stack plot (Figure 5(c)),
the mass flows started at about 14:20 UT and continued in the
gradual phase of the flare. The duration of mass flows was more than
40 min and the projected velocity in the plane of the sky was 50-75
km s$^{-1}$. The flows were generated intermittently and showed a
quasi-periodic pattern with an average period of about 90 s (brown
cure in Figure 5(c)). In addition, at about 14:25 UT, a secondary
flare ribbon SR (Polito et al. 2016) appeared between ribbons R1 and
R2 and elongated towards ribbon R2 at a speed of 90 km s$^{-1}$
(Figures 5(a)-(c)). From the 304 {\AA} stack plot (Figure 5(d)), the
long-duration flows can also be clearly displayed, which moved at a
speed of 60-110 km s$^{-1}$.

The \emph{IRIS} spectral observations of Si {\sc iv} 1402.77 {\AA}
line cover the east part of the filament and the
reconnection-induced flows. The locations of Slit 0, Slit 4 and Slit
7 are focused on (Figures 6(a) and (d)) and their spectral images of
the Si {\sc iv} 1402.77 {\AA} line at two different times are
displayed in Figures 6(b1)-(b3) and (e1)-(e3). The Si {\sc iv} line
exhibited evident blueshifts at the locations of the flows along the
three slits. At 14:28:57 UT, the spectral tilt was clearly visible
at the location of Slit 0 (Figure 6(b1)), suggestive of the spinning
motion along the magnetic structure of the filament (Curdt et al.
2012; Li et al. 2014). The 1330 {\AA} SJIs show that the top edge of
the intersection area between the slit locations and the filament
seems to be along the same fine structure of the filament (FS in
Figures 6(a) and (d)). Three different locations ``A", ``B" and ``C"
along FS are selected, and their spectral profiles and corresponding
single-Gaussian fits are shown in Figures 6(c) and (f). For the
location ``A", the calculated blueshift reached about 121 km
s$^{-1}$ and the non-thermal width was 61 km s$^{-1}$ at 14:28:57 UT
(Figure 6(c)). For the locations closer to the reconnection region
(``B" and ``C"), the blueshifts decreased and the non-thermal widths
remained the same. The blueshift at location ``B" decreased to 82 km
s$^{-1}$ and location ``C" at the west had a blueshift of only 64 km
s$^{-1}$. At 14:38:17 UT, the blueshift at location ``A" increased
to 143 km s$^{-1}$ with a large non-thermal width of about 150 km
s$^{-1}$ (Figure 6(f)). Similarly, the Doppler shifts at the
locations near the reconnection site are smaller than those
locations far away. The difference of Doppler shifts along the same
fine structure probably implies the helical magnetic field of the
filament.

In order to analyze the temporal evolution of Doppler shifts and
non-thermal widths at the reconnection-induced flows, we fit the Si
{\sc iv} 1402.77 {\AA} profiles and obtained the stack plots of peak
intensity, Doppler shift and non-thermal width along Slit 0 and Slit
4 (Figure 7). Ribbon R2 showed an evident separation motion with a
speed of about 18 km s$^{-1}$, however, ribbon R1 located in the
sunspot did not show significant displacement perpendicular to the
ribbon (Figures 7(a) and (d)). The Si {\sc iv} emission exhibited
obvious redshifts of 20-90 km s$^{-1}$ at ribbon R2 (Figures 7(b)
and (e)), suggestive of the chromospheric condensation during the
process of explosive chromospheric evaporation (Fisher et al. 1985;
Zhang et al. 2016; Li et al. 2017). At ribbon R1, the Si {\sc iv}
line was blueshifted by about 10-20 km s$^{-1}$, probably due to the
combination effect of the chromospheric condensation and the
reconnection-induced flows at the footpoints of the filament. About
8 min after the flare onset, the Doppler blueshifts and broadening
at the location of flows appeared, with the blueshift of 50-240 km
s$^{-1}$ (Figures 7(b) and (e)) and the non-thermal width reaching
about 60-120 km s$^{-1}$ (Figures 7(c) and (f)). The blueshift
lasted for more than 40 min and can be observed throughout the
impulsive and gradual phases of the flare.

\section{Summary and Discussion}

We have presented an extensive analysis of an X2.0-class confined
flare in AR 12192 on 2014 October 27. The flare mainly consisted of
two sets of magnetic systems, with their east ends anchoring in the
same negative-polarity ribbon (R1) above a sunspot umbra. The west
ends of the two magnetic systems were respectively located in other
two positive-polarity ribbons (R2 and R3). Using the \emph{IRIS}
high-resolution observations, we report the apparent slipping motion
of ribbon substructures along ribbon R1 in the TR. The majority of
the substructures exhibited the slipping motion towards the hook of
ribbon R1 away from the inversion line and a small fraction was seen
to slip in the opposite direction towards the straight part of R1,
with apparent speeds of 10-70 km s$^{-1}$. At the same location, the
substructures intermittently appeared with a period of 80-110 s,
indicating the quasi-periodic pattern of the slippage. The slipping
motion can be observed since the flare start and throughout the
impulsive phase of the flare. A reverse-S shaped filament connecting
ribbons R1 and R2 was activated when the slippage along ribbon R1
reached the end of the filament. Simultaneously continual
reconnection-induced flows moved along the filament from its middle
part to the east end, with a speed of 50-110 km s$^{-1}$ in the
plane of the sky. The flows were generated quasi-periodically with
an average period of about 90 s. The period of the flows is similar
to that of the slipping motion of ribbon substructures, implying the
occurrence of quasi-periodic slipping magnetic reconnection. The
spectral observations showed that at the reconnection-induced flows
the Si {\sc iv} line was blueshifted by 50-240 km s$^{-1}$ and
broadened with the non-thermal width reaching about 60-120 km
s$^{-1}$. The Doppler blueshifts lasted for about 40 min and can be
observed until the gradual phase of the flare. During the process of
the flare, the filament did not show evident rise phase and did not
erupt at all.

During this flare, the magnetic field line linkage has two basic
sets of magnetic connectivities (S1 and S2), with their east ends
both anchoring in negative-polarity ribbon R1 and their respective
west ends in two positive-polarity ribbons R2 and R3. The
bi-directional slipping motion of ribbon R1 and the simultaneous
elongations of ribbons R2 and R3 imply the existence of a QSL
surface between the two magnetic systems (S1 and S2). We suggest
that the magnetic reconnection occurs along the QSL and causes the
continuous exchange of connectivities of two magnetic systems S1 and
S2, which results in apparent slippage of the entire reconnecting
field lines. The QSL reconnection also generates the observed flows
along the non-eruptive filament. According to the connectivity of
the filament, we suggest that the filament belongs to system S1.
When the slippage of R1 reached the eastern end of the filament, the
slipping reconnection between field lines of the filament and system
S2 occurred along the QSL and caused the quasi-periodic flows along
the filament. A relevant simulation study was carried out by
Aulanier et al. (2005), who performed the MHD simulations of the
development of electric currents in bipolar configurations. In that
study, two sets of magnetic connectivities appeared in the bipolar
potential configurations and slipping magnetic reconnection occurred
along the QSL between the two systems (see their Figure 3), similar
to the magnetic topology in our observations.

The bi-directional slippage of the substructures within the same
flare ribbon in the TR is first reported in this study. We suggest
that the slipping substructures in two directions respectively
correspond to the footpoints of two magnetic systems. The continuous
slipping magnetic reconnection between the two magnetic systems
results in the exchange of their connectivities and the slippage of
reconnecting field lines in opposite directions. Using the
\emph{SDO}/AIA observations, Dud{\'{\i}}k et al. (2016) reported the
apparent slipping motions of flare loops in opposite directions
during an eruptive flare. They interpreted the slippage towards the
ribbon hook corresponding to the field lines of the erupting flux
rope and the slippage in the opposite direction corresponding to the
flare loops, fullfilling the prediction of the standard solar flare
model in 3D. Different from their observations, there is no erupting
flux rope involved in the analyzed confined flare and our
observations can not be explained by the 3D standard flare model.
Similar to the behavior of ribbons substructures in the TR, the
chromospheric knots within a ribbon-like structure have also been
observed to move in two directions towards both ends of the ribbon
(Sobotka et al. 2016). These observations imply that the energy
release by slipping magnetic reconnection can affect both the TR and
the chromosphere.

The quasi-periodic property of slipping kernels within flare ribbons
has been revealed during an eruptive flare by Li \& Zhang (2015),
and was further verified by Brosius \& Daw (2015) and Brannon et al.
(2015). All these observations about the sawtooth oscillation of
ribbon substructures are for eruptive flares, and the reported
period range is between 90 s and 300 s. In our observations, the
small-scale substructures ($\sim$500 km) within flare ribbons also
exhibit the quasi-periodic slipping motion in the confined flare,
and the oscillation period of 80-110 s is similar to that during
eruptive flares. Recently, Parker \& Longcope (2017) proposed a
model of the tearing mode instability and reproduced the
observational signatures of the sawtooth oscillation. They suggested
that a tearing mode with asymmetric shear flow occurring at a
current sheet produced fluid motions similar to the observations.
Another explanation about the sawtooth oscillation of ribbon
substructures is related to the modulation of MHD waves (McLaughlin
et al. 2009; Nakariakov \& Melnikov 2009). The MHD waves during the
flare may affect the reconnection site and cause the oscillatory
magnetic reconnection.

The simultaneous imaging and spectroscopic observations of the
reconnection-induced flows in the confined flare are presented in
this work. The flows moved along the QSLs at a speed of 50-240 km
s$^{-1}$ in the impulsive and gradual phases of the flare. Here, the
flows should not be confused with the downflows along post-flare
loops due to the material cooling after magnetic reconnections
(Schmieder et al. 1996; Song et al. 2016). The appearance of
reconnection-induced flows in confined flares differ from that in
eruptive flares. In eruptive flares, bi-directional outflows are
usually in forms of plasmoid ejections and contracting cusp-shaped
loops at speeds of about 100-300 km s$^{-1}$ (Savage \& McKenzie
2011; Liu et al. 2013), comparable to the Doppler shifts of
reconnection-induced flows in our event. Moreover, the filament
involved in the flare seemed to take part in the QSL reconnection,
however, was still present after the flare. It did not show the rise
phase and was not associated with any failed eruption. The
appearance of the filament and the reconnection-induced flows could
not be explained by the classical picture of confined flares, either
due to a failed filament eruption (Ji et al. 2003; Kushwaha et al.
2015) or the loop-loop interaction (Green et al. 2002; Ning et al.
2018). Thus the physical factors determining the likelihood of
ejective/confined eruptions need to be revisited. We suggest that
the QSL reconnection in complex magnetic configuration caused no
significant topological change, and thus the stability of the
filament was not destroyed and the flare eventually developed into a
confined event.

\acknowledgments {We thank Miho Janvier for useful discussions.
\emph{SDO} is a mission of NASA's Living With a Star Program.
\emph{IRIS} is a NASA small explorer mission developed and operated
by LMSAL with mission operations executed at NASA's Ames Research
center and major contributions to downlink communications funded by
the Norwegian Space Center (NSC, Norway) through an ESA PRODEX
contract. This work is supported by the National Natural Science
Foundations of China (11773039, 11533008, 11790304, 11673035,
11673034 and 11790300), Key Programs of the Chinese Academy of
Sciences (QYZDJ-SSW-SLH050), and the Youth Innovation Promotion
Association of CAS (2017078 and 2014043).}

{}
\clearpage

\begin{figure}
\centering
\includegraphics
[bb=27 223 534 594,clip,angle=0,scale=0.9]{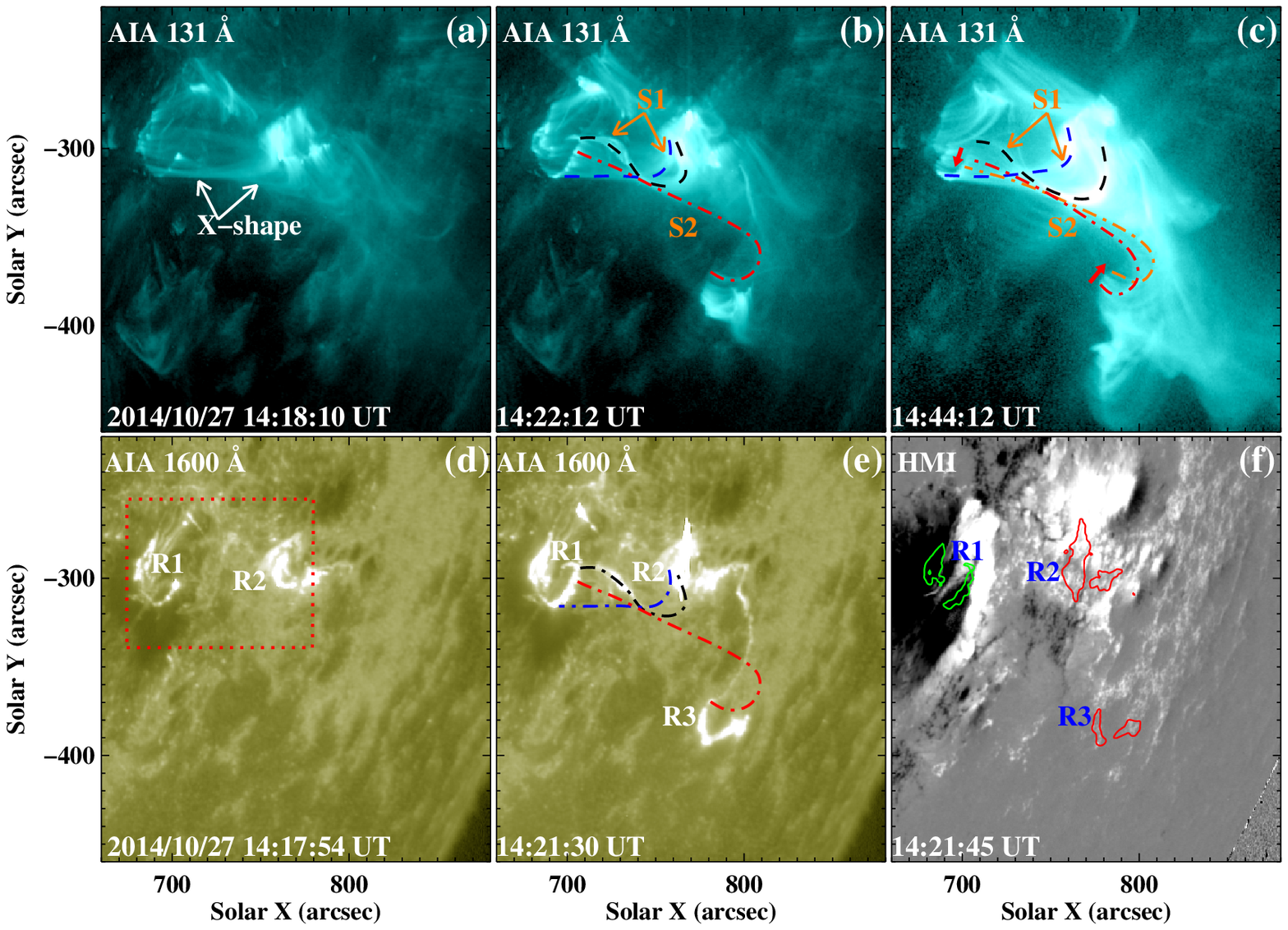}
\caption{Overview of the confined X2.0 flare on 2014 October 27 in
AR 12192 including the multi-wavelength images from \emph{SDO}/AIA
and \emph{SDO}/HMI LOS magnetogram. The structure S1 denotes the
``X-shaped" flare loops outlined by black and blue dashed curves.
The red and brown dash-dotted curves show another set of magnetic
structure S2 involved in the flare. The red arrows in panel (c)
represent the two opposite slipping directions of the footpoints of
structure S2. The duplications of structures S1 and S2 are also
shown in panel (e). The red rectangle in panel (d) shows the FOV of
the images in Figure 2. R1, R2 and R3 are three flare ribbons, and
their brightness contours in AIA 1600 {\AA} image are shown in panel
(f). In the associated animation, the AIA 131 {\AA} and AIA 1600
{\AA} image animations start at 14:05:20 UT and 14:05:04 UT,
respectively. They end at 15:29:47 UT and 15:29:28 UT. The HMI
magnetograms are not included in the animation since the evolution
of magnetic fields is not evident during this period. \label{fig1}}
\end{figure}
\clearpage

\begin{figure}
\centering
\includegraphics
[bb=23 170 543 651,clip,angle=0,scale=0.85]{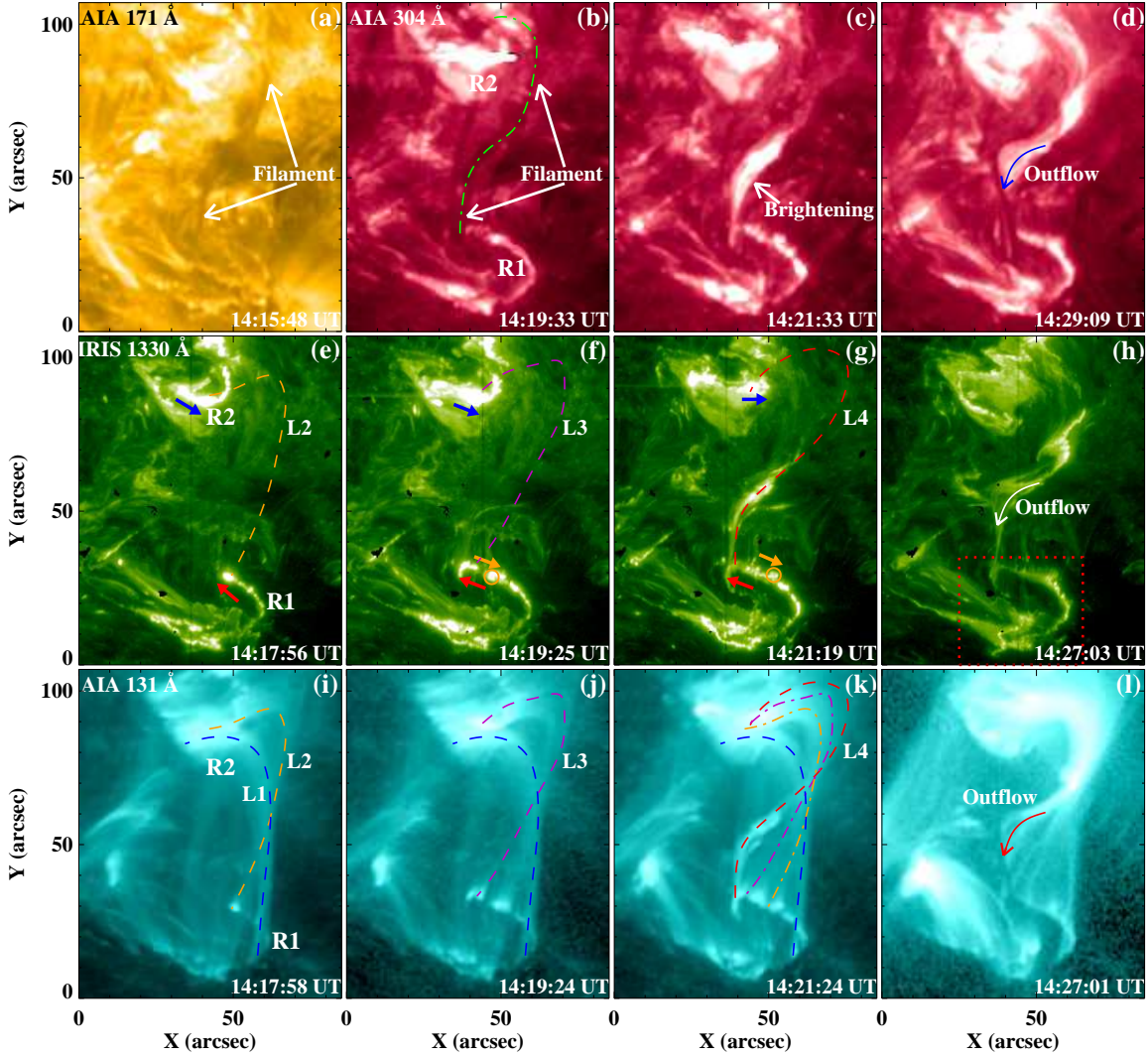}
\caption{Multi-wavelength (E)UV images from the \emph{SDO}/AIA and
\emph{IRIS} showing the dynamic evolution of the flare. The AIA
images are rotated by $90^{\circ}$ counterclockwise to co-align with
the \emph{IRIS} SJIs. The green dash-dotted curve in panel (b)
outlines the filament involved in the event. The red arrows in
panels (e)-(g) show the slippage of ribbon R1 towards the north, and
the brown circles and arrows denote one of moving substructures
within ribbon R1 and its opposite slipping direction. Blue arrows in
panels (e)-(g) denote the elongation direction of ribbon R2. Dashed
lines in panels (i)-(k) outline the high-temperature flare loops
connecting ribbons R1 and R2, and dash-dotted lines are the
duplicates of the flare loops at earlier times. The duplications of
loops L2-L4 are also shown in panels (e)-(g). The red rectangle in
panel (h) shows the FOV of the 1330 {\AA} images in Figure 3. The
associated animation includes \emph{IRIS} 1330 {\AA}, AIA 304 {\AA},
171 {\AA} and 131 {\AA} images from about 14:12 UT to 15:00 UT.
\label{fig2}}
\end{figure}
\clearpage

\begin{figure}
\centering
\includegraphics
[bb=26 185 541 632,clip,angle=0,scale=0.85]{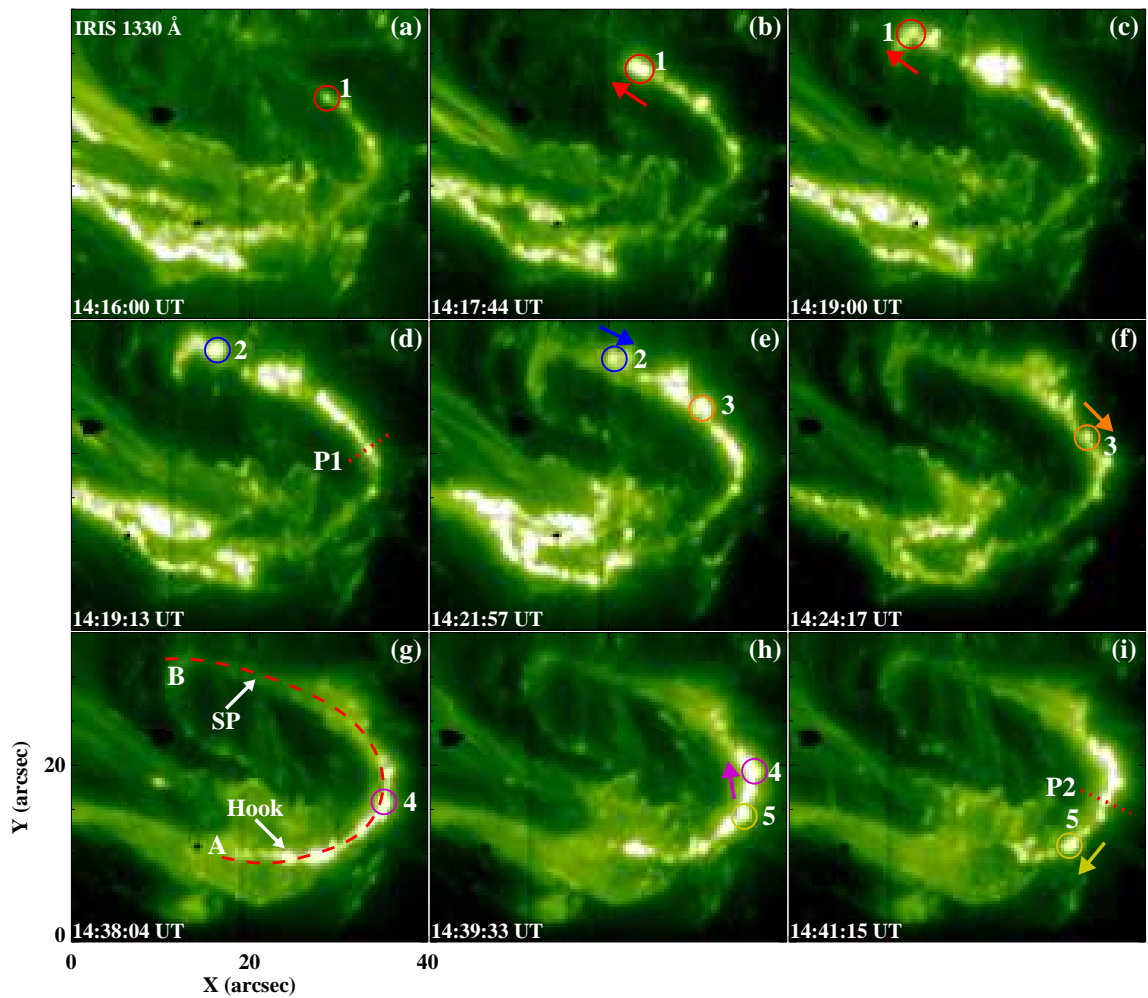} \caption{Time
series of 1330 {\AA} images showing the bi-directional slippage of
the substructures within ribbon R1. The circles with different
colors show five of slipping substructures (``1"$-$``5"), and the
arrows point to the slipping directions. Dotted lines ``P1" in panel
(d) and ``P2" in panel (i) denote the two cuts shown in Figures
4(c)-(d). Dashed curve ``A-B" (panel (g)) represents the location
used to obtain the stack plot shown in Figure 4(a). ``A" is at the
hook part of the ribbon and ``B" is at the straight part (SP) near
the inversion line. The associated animation includes the zoomed
\emph{IRIS} 1330 {\AA} images, starting at 14:11:59 UT and ending at
15:04:49 UT. \label{fig3}}
\end{figure}
\clearpage

\begin{figure}
\centering
\includegraphics
[bb=23 231 572 542,clip,angle=0,scale=0.85]{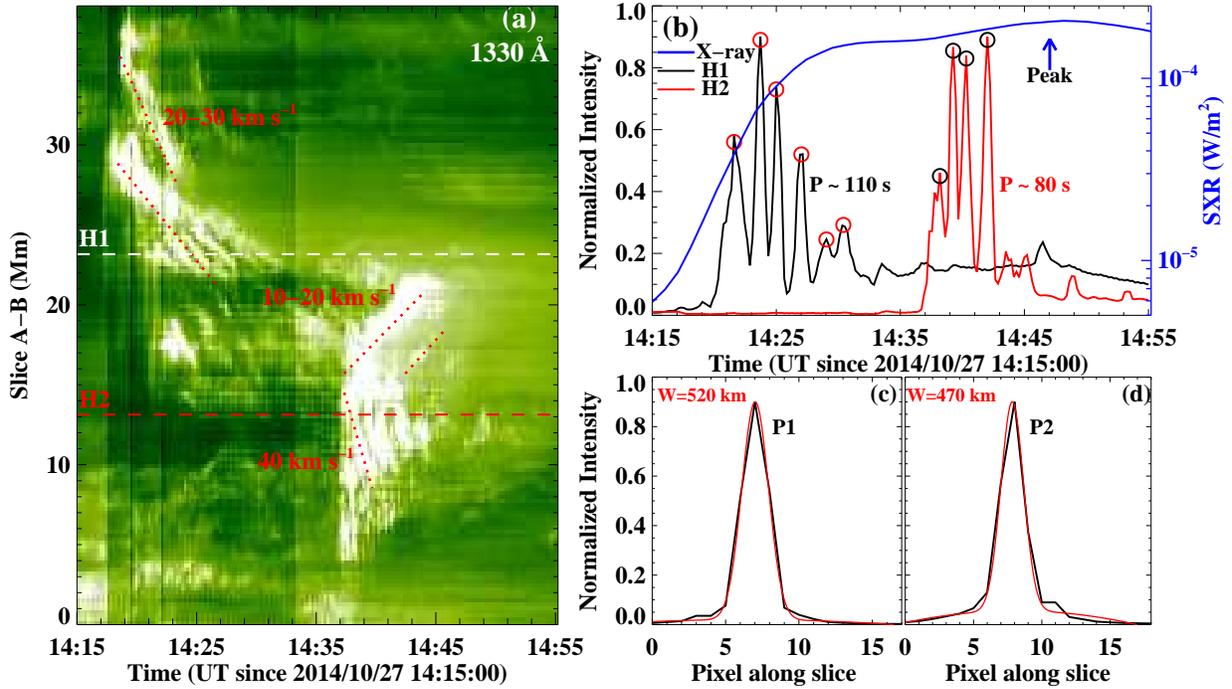} \caption{Panel
(a): 1330 {\AA} stack plot along slice ``A-B" (Figure 3(g))
displaying the quasi-periodic slipping motion of ribbon
substructures. Two horizontal lines ``H1" and ``H2" denote the
locations where the horizontal slices in panel (b) are obtained.
Panel (b): GOES SXR 1-8 {\AA} flux (blue curve) of the flare and
horizontal slices (black and red curves) along straight lines ``H1"
and ``H2". Panels (c)-(d): intensity-location profiles (black
curves) and their Gaussian fitting profiles (red curves) along the
two cuts ``P1" and ``P2" (Figures 3(d) and (i)) showing the widths
(W) of the substructures. \label{fig4}}
\end{figure}
\clearpage

\begin{figure}
\centering
\includegraphics
[bb=101 160 462 721,clip,angle=0,scale=0.9]{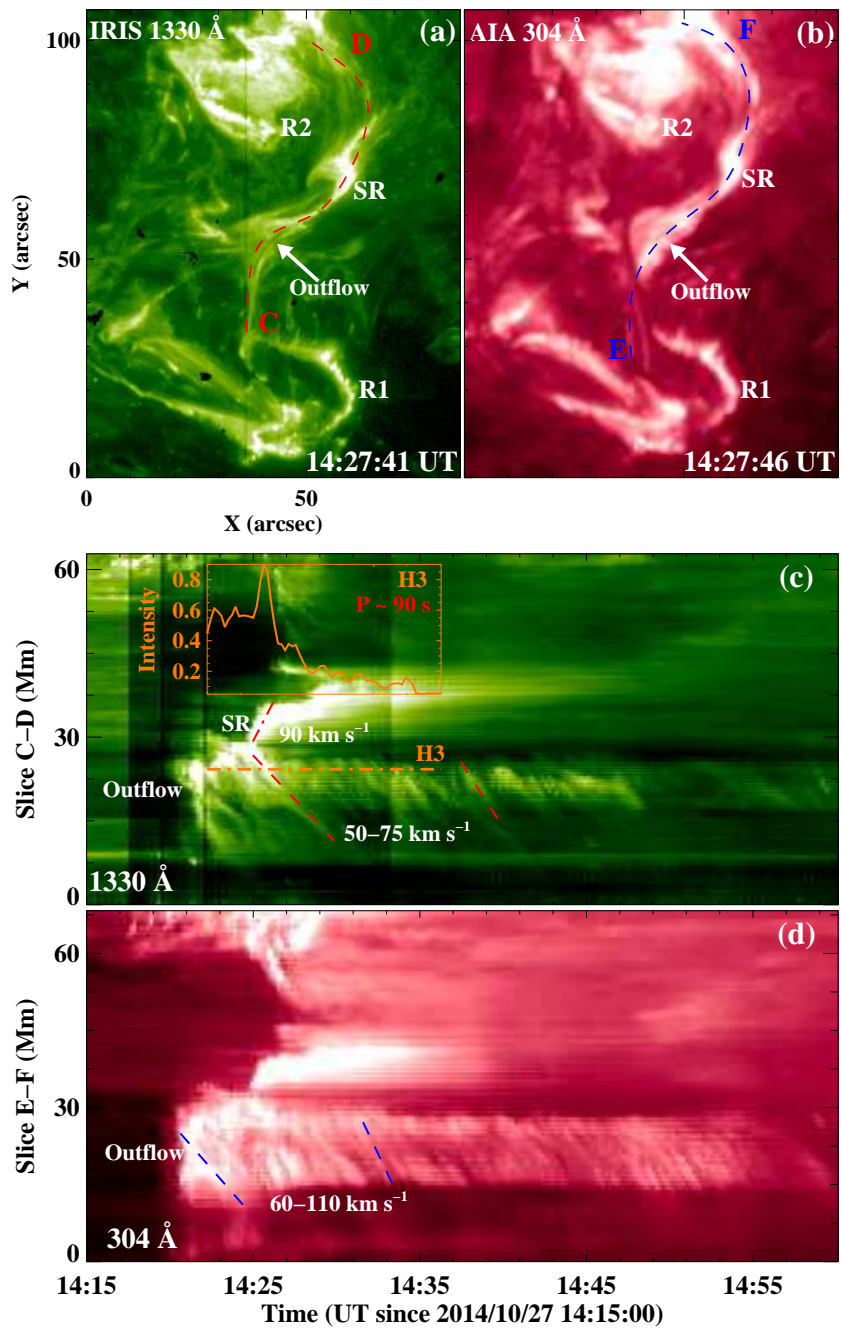}
\caption{Continual reconnection-induced flows along the non-eruptive
filament in 1330 and 304 {\AA} observations. The FOV of panels
(a)-(b) is the same as the images in Figure 2. The red and blue
curves (``C-D" and ``E-F") in panels (a)-(b) respectively denote the
locations where the two stack plots in panels (c)-(d) are obtained.
SR in panels (a)-(c) represents the secondary ribbon between the two
ribbons R1 and R2. The overplotted brown curve in panel (c) shows
the normalized intensity along the horizontal brown dash-dotted line
``H3". \label{fig5}}
\end{figure}
\clearpage

\begin{figure}
\centering
\includegraphics
[bb=17 178 546 642,clip,angle=0,scale=0.9]{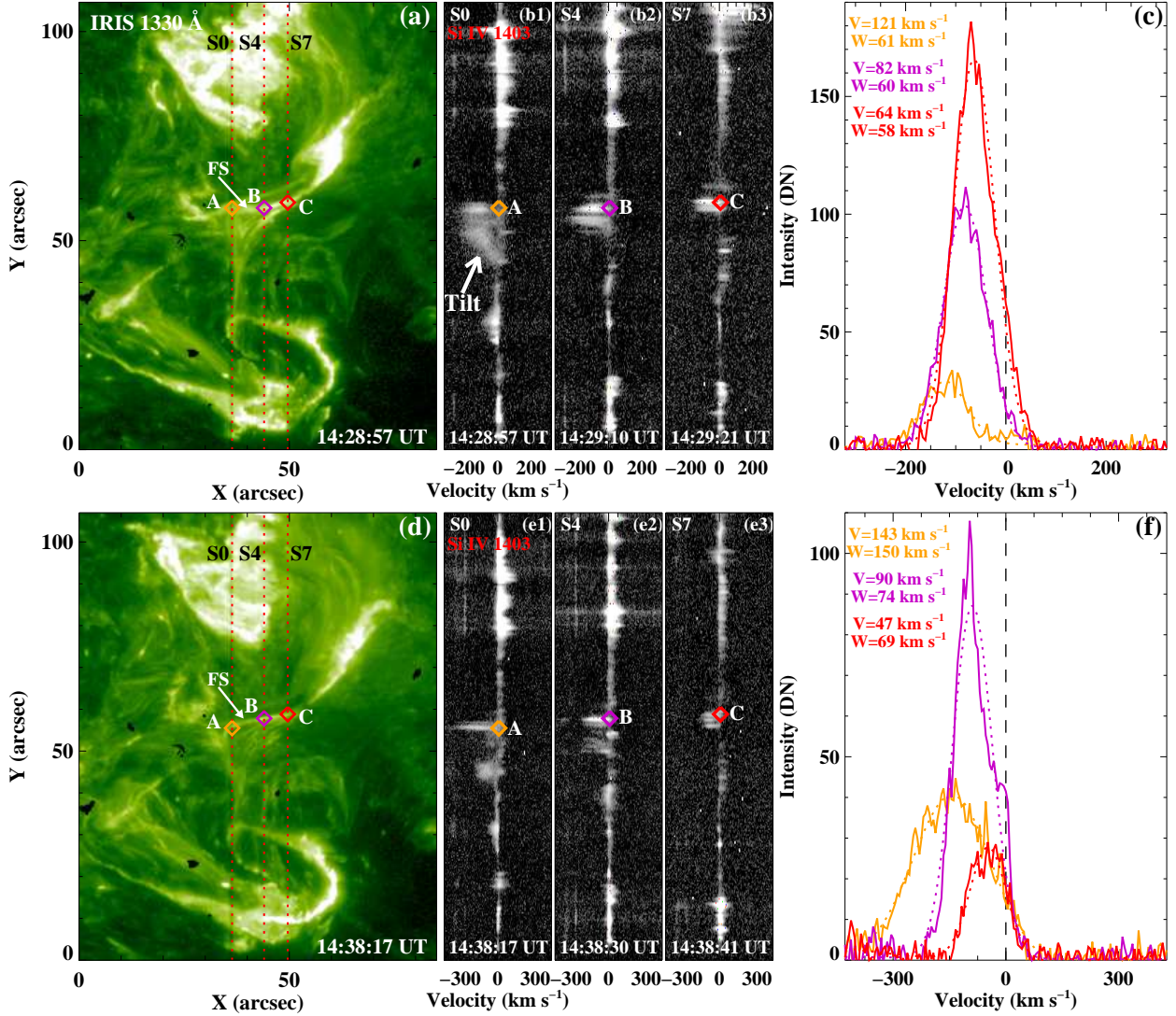} \caption{Left
column: \emph{IRIS} 1330 {\AA} SJI images displaying the slit
locations and the fine structure of the filament (FS). Their FOV is
the same as Figure 2. The red dotted lines denote three slit
locations S0, S4 and S7, and the diamonds (``A", ``B" and ``C") show
three selected locations along the structure FS of the filament.
Middle column: spectral images of the Si {\sc iv} 1402.77 {\AA} at
the three slits as shown in panels (a) and (d). Right column:
profiles of the Si {\sc iv} line at locations ``A", ``B" and ``C".
The solid curves are the observed profiles and the dotted curves are
single-Gaussian fitting profiles. \label{fig6}}
\end{figure}
\clearpage

\begin{figure}
\centering
\includegraphics
[bb=45 176 517 684,clip,angle=0,scale=0.9]{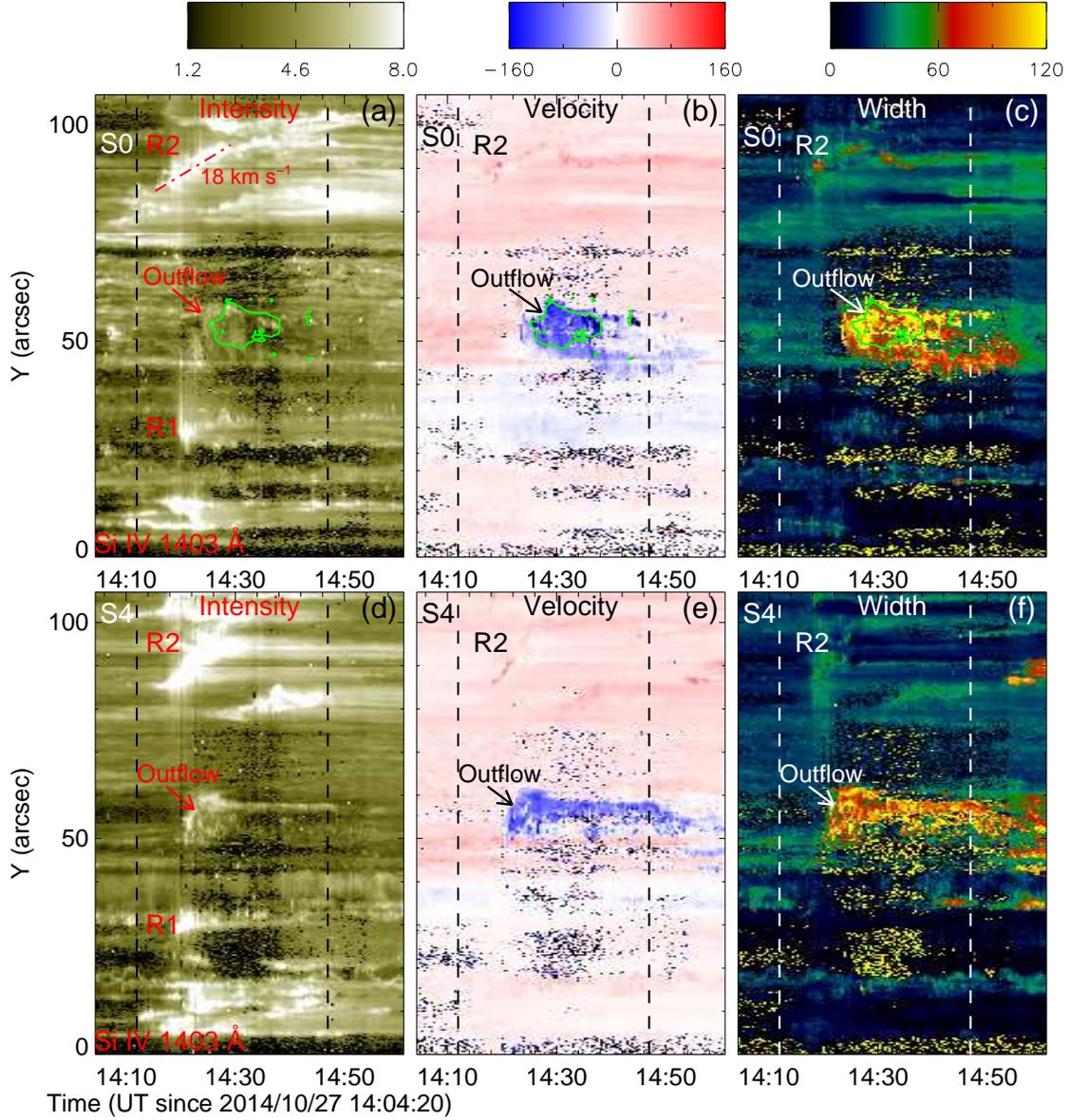} \caption{Top
row: temporal evolution of peak intensity, Doppler shift and
non-thermal width through fitting the Si {\sc iv} profiles along S0.
The spatial range is along the Y axis in Figures 6(a) and (d). The
green contours are the Doppler blueshifts at the level of 80 km
s$^{-1}$. Two dashed lines respectively denote the flare start
(14:12 UT) and peak time (14:47 UT). Bottom row: the same as the top
row but for S4. \label{fig7}}
\end{figure}
\clearpage

\end{document}